\documentstyle[preprint,aps,psfig]{revtex}
\begin{document}
\tighten
%
% uncomment \draft to have PACS numbers appear
\draft
% put preprint number(s).
%\preprint{}
\title{Strangeness-changing response functions: \\
       an alternative approach to hypernuclear structure}
\author{H.~M\"uller}
\address{Department of Physics, University of Colorado,
Boulder, Colorado 80309}
\author{J.~Piekarewicz}
\address{Department of Physics and Supercomputer Computations
	 Research Institute, \\ Florida State University,
         Tallahassee, Florida 32306-4130}
\vskip1in
%\date{\bf DRAFT: \today}
\date{\today}
\maketitle
\begin{abstract}
{We study strangeness-changing response functions as an alternative
description of hypernuclear structure. Hypernuclear states are treated
in the same manner as any other conventional nuclear excitation that
emerges from the response of the nuclear ground state to an external
probe. The nuclear response is calculated using a random-phase
approximation to an effective relativistic mean-field model that
reproduces bulk properties of normal nuclei. The relevant meson-baryon
vertices are constrained by imposing SU(3)-flavor symmetry, while the
residual hyperon-particle---nucleon-hole interaction is assumed to be
mediated by the two lightest $S\!=\!-1$ mesons: the pseudoscalar kaon
and its vector partner the $K^*(892)$. We use this model to analyze
the spectra of $^{16}_{Y}$O and $^{40}_{Y}$Ca hypernuclei.}
\end{abstract}
\vspace{20pt}
%
%###########################################################################%
%###########################################################################%
% This is file chap1.tex
%%%%%%%%%%%%%%%%%%%%%%%%%%%
%
\section{Introduction}
\label{intro}
Strangeness adds yet another --- still largely unexplored ---
dimension to hadronic physics. For nucleons in free space, their
strange-quark content has received considerable attention as a result
of the measurement of the spin-dependent structure function of the
proton by the European Muon Collaboration (EMC)~\cite{ashman88}.  Some
analyses of the experiment suggest that a large portion of the spin of
the proton is carried by strange quarks. In the opposite,
high-density, domain a copious production of strangeness is predicted
as soon as the Fermi energy of the system becomes large enough for the
addition of strange quarks to become energetically
favorable~\cite{review98}.  Thus, one expects a substantial increase
in the strangeness-per-baryon ratio with density.  In addition to its
obvious relevance to nuclear physics, high-density matter appeals to
many other branches of physics, such as astrophysics, cosmology and
particle physics.  Indeed, the high-density environment found at the
core of neutron stars is speculated to constitute a fruitful ground
for the formation of exotic states of matter --- such as a kaon
condensate and strange-quark matter. Motivated by these fundamental
issues a large number of experimental facilities are devoting valuable
resources to elucidate them. However, before a clear picture of these
exotic phenomena will emerge, it is necessary to understand phenomena
at or near normal nuclear-matter density. The formation of hypernuclei
appears to be an important first step. Moreover, because
hyperon-nucleon scattering experiments are difficult to perform,
hypernuclear physics appears to be an attractive alternative for
testing --- and for extending off the mass shell --- models of the
elementary nucleon-hyperon and hyperon-hyperon interaction. Indeed, a
vast number of experiments have been devoted~\cite{BANDO90} to study
the formation, the spectroscopy and the decay of hypernuclei.

To date, hypernuclei have been typically produced through
strangeness-exchange reactions using hadronic probes --- such as in
$(K^-,\pi^-)$ and $(\pi^+,K^+)$ scattering --- and decay through
nonleptonic weak processes which involve the emission of nucleons
and pions. These experiments have been analyzed from a variety of
different theoretical perspectives. For example, microscopic
meson-exchange models have been constructed which accurately reproduce
the rich nucleon-nucleon and the more scarce hyperon-nucleon
data~\cite{NAGELS78,HOLZENKAMP89}.  Other approaches, more closely
connected to the underlying symmetries of QCD, are based on effective
field theories that include the baryon octet and several nonstrange
and strange mesons~\cite{SCHAFFNER93,PAPAZOGLOU98a,MUELLER99}. With
such models available, binding energies and single-particle spectra of
hypernuclei have been computed in the context of
nonrelativistic~\cite{RAYET81} and relativistic mean-field models
\cite{RUFA87}. The underlying assumption in the mean-field approach
is that the dynamical aspects of the production and decay process can
be ignored, so that the hypernucleus may be approximated as a core of
nucleons with one strange baryon added to the system. If the hyperon
has been produced from a closed-shell nucleus this approximation
constitutes the simple particle-hole model. Yet one must take this
simplified description with caution as the real picture is vastly
more complicated. Indeed, it is widely recognized that the nuclear
resonances populated through these reactions are often highly excited
(particle-hole) states that may decay electromagnetically prior to
their weak decay. Moreover, initial- and final-state interactions
between the nucleus and the hadronic probes are strong and must be
included. These aspects add considerable complexity to any theoretical
description. Although the gross features of hypernuclear spectra may
indeed be explained on the basis of a simple particle-hole picture,
the fine details require the incorporation of distortions as well as
of few- and many-body correlation effects~\cite{BRUECKNER76}.

With the advent of continuous electron beam facilities, such as the
Thomas Jefferson National Accelerator Facility (TJNAF), the formation
of hypernuclear states through the photoproduction of charged kaons
has become a reality ~\cite{BERNSTEIN81}.  The production of
hypernuclei via hadronic or photo-nuclear reactions is a complicated
process involving not only nucleons and hyperons but also several
baryon resonances and strange mesons.  In contrast to the hadronic
production, however, the $K^{+}$-photoproduction process is relatively
insensitive to distortion effects. This represents an enormous
theoretical simplification. Still, additional simplifications and
approximations are often needed. For example, in order to keep the
influence of the nuclear environment on a tractable level, most
theoretical studies resort to the impulse approximation: the
assumption that the elementary process suffers no modification as it
is embedded in the nuclear medium. (Note that the free-space
production operator could suffer considerably from model and parameter
dependencies~\cite{COHEN89}.)  Further, if distortion effects can be
neglected --- an excellent approximation in the
$K^{+}$-photoproduction case --- then as in the case of electron
scattering, the theoretical amplitude may be decomposed into a nuclear
structure part and a production
operator~\cite{COHEN89,BENNHOLD89,HUEFNER74,DOVER80}. The production
operator contains the specific details of the reaction, including
various meson-baryon vertices and form factors. It is obtained from
the corresponding $T$-matrix describing the reaction on a single
nucleon in free space. The nuclear structure part involves the
response of the nuclear ground state in the form of several
polarization insertions describing the propagation of a nucleon-hole
and a $\Lambda$($\Sigma^0$)-particle through the nuclear medium.

Using this point of view, we study hypernuclear spectra using a
random-phase approximation (RPA) to a generalized relativistic
mean-field model \cite{FST97}. More specifically, baryonic matter is
assumed to consist of strange as well as non-strange hadrons. At, or
near, nuclear-matter saturation density --- and for phenomenologically
consistent hyperon-meson couplings --- the presence of hyperons in the
ground state is suppressed by their larger mass. Thus, at these
densities one recovers the normal [strangeness-equal-zero ($S=0$)]
nuclear ground state. Hypernuclear states, as well as any other
nuclear excitation, then emerge from the consistent response of this
mean-field ground state to a suitable external probe. That is, we
propose to treat strangeness no different than any other intrinsic
nuclear quantum number. Moreover, as the RPA is the consistent
response of the mean-field ground state, all nuclear states --- those
with $S=0$ and $S\ne 0$ alike --- are obtained by correlating the
particle-hole pair via their residual interaction. The importance of
many-body correlations beyond the simple particle-hole picture was
realized soon after precise hypernuclear measurements were
available~\cite{BRUECKNER76}.  For example, mixing of particle-hole
states that couple to the same angular momentum is essential for
predicting angular distributions in strangeness exchange
reactions~\cite{CHIANG79}.  Furthermore, states with nucleon holes in
the deepest bound nucleon shells are not recognized in the experiments
because of their large widths, a feature which cannot be explained by
simple mean-field models~\cite{HUEFNER74b}.

The outline of this paper is as follows: In Sec.~\ref{lagrangian}, we
present the effective Lagrangian which contains the relevant
baryon-meson vertices.  Section~\ref{response} is devoted to a brief
discussion of strangeness-changing response functions.
We also discuss the residual particle-hole interaction
and the RPA equations. Section~\ref{results} contains
specific results for hypernuclear spectra.  Finally,
Sec.~\ref{summary} contains a short summary.
%
% This is file chap2.tex
%%%%%%%%%%%%%%%%%%%%%%%%%%%
%
\section{The effective Lagrangian}
\label{lagrangian}
To describe the nuclear ground state we use a relativistic
mean-filed model based
on an effective Lagrangian that realizes chiral
symmetry and vector meson dominance.
In the nucleon sector this approach has been successful in describing the
ground state properties of ordinary nuclei \cite{FST97}.
More recently, these ideas have been generalized to include strangeness
\cite{MUELLER99}.

The Lagrangian is formulated in terms of the effective degrees of 
freedom that are taken to be the baryon octet, the Goldstone boson 
octet and the vector meson nonet:
\begin{equation}
{\cal L}= {\cal L}_{BM}+{\cal L}_M + {\cal L}_{EM} \ ,
\label{eq:l0}\\
\end{equation}
where the subscript $M (MB)$ denotes the meson (meson-baryon) sector
and $EM$ the electromagnetic interaction.
The octets are collected in $3\times3$ 
traceless hermitian matrices
\begin{equation}
B = \left(\begin{array}{ccc}
\frac{1}{\sqrt{6}}\Lambda+\frac{1}{\sqrt{2}}\Sigma^0 & \Sigma^+  & p \\
\Sigma^- & \frac{1}{\sqrt{6}}\Lambda-\frac{1}{\sqrt{2}}\Sigma^0 &  n \\
\Xi^- & \Xi^0& -\frac{2}{\sqrt{6}}\Lambda\\
\end{array} \right) \ ,
\label{eq:baryon}
\end{equation}
\begin{equation}
\Pi = \left(\begin{array}{ccc}
\frac{1}{\sqrt{6}}\eta+\frac{1}{\sqrt{2}}\pi^0 & \pi^+  & K^+ \\
\pi^- & \frac{1}{\sqrt{6}}\eta-\frac{1}{\sqrt{2}}\pi^0 &  K^0 \\
K^- & \overline{K}^0& -\frac{2}{\sqrt{6}}\eta\\
\end{array} \right) \ ,
\label{eq:gb}
\end{equation}
\begin{equation}
V_\mu = \left(\begin{array}{ccc}
\frac{1}{\sqrt{6}}V^8_\mu+\frac{1}{\sqrt{2}}\rho^0_\mu & \rho^+_\mu 
& K^{*+}_\mu \\
\rho^-_\mu & \frac{1}{\sqrt{6}}V^8_\mu-\frac{1}{\sqrt{2}}\rho^0_\mu 
&  K^{*0}_\mu \\
K^{*-}_\mu & \overline{K}^{*0}_\mu& -\frac{2}{\sqrt{6}}V^8_\mu\\
\end{array} \right) \ .
\label{eq:vector}
\end{equation}
The physical $\omega$ and $\phi$ mesons arise 
from the mixing of the $V^8_\mu$ and the vector meson singlet $S_\mu$ via
\begin{eqnarray}
\omega_{\mu}&=& \cos(\theta) S_{\mu} + \sin(\theta) V^8_{\mu}  \ ,
\label{eq:mesonmix} \\
\phi_{\mu}&=& \sin(\theta) S_{\mu} - \cos(\theta) V^8_{\mu}  \ .
\nonumber
\end{eqnarray}
We also include a light isoscalar scalar meson $\varphi$ 
which simulates the exchange of correlated pions and kaons.

The couplings of the mesons to the baryons are contained in
\begin{eqnarray}
{\cal L}_{MB}^{\prime} &=& 
  F{\rm Tr}\Bigl({\overline B}i\gamma_5[\Delta\!\!\!/, B]\Bigr)
+ D{\rm Tr}\Bigl({\overline B}i\gamma_5\{\Delta\!\!\!/, B\}\Bigr)
\label{eq:lmb}\\
&&\null- g_F {\rm Tr}\Bigl({\overline B}[V\!\!\!\!/, B]\Bigr)
        - g_D {\rm Tr}\Bigl({\overline B}\{V\!\!\!\!/, B\}\Bigr)
        - g_S {\rm Tr}\Bigl({\overline B}S\!\!\!/B\Bigr) \nonumber \\
&&\null - {f_F\over 4M} {\rm Tr}\Bigl({\overline B}
          [\sigma_{\mu\nu}V^{\mu\nu}, B]\Bigr)
        - {f_D \over 4M}{\rm Tr}\Bigl({\overline B}
          \{\sigma_{\mu\nu}V^{\mu\nu}, B\}\Bigr)
        - {f_S\over 4M} {\rm Tr}\Bigl({\overline B}
          \sigma_{\mu\nu}S^{\mu\nu} B\Bigr) \ ,
\nonumber
\end{eqnarray}
where we disregard terms generated by the covariant derivative of the baryons.
The pseudo-vector couplings of the kaons to the baryons arise from 
the expansion
\begin{eqnarray}
\Delta_{\mu}&=&\frac{1}{2}(u^{\dagger}\partial_{\mu}u -
u\partial_{\mu}u^{\dagger}) = {i\over \sqrt{2}f} \partial_\mu \Pi + \ldots
\label{eq:delta} \\
&&\null \hbox{\rm for}\quad u=e^{{i\over \sqrt{2}f}\Pi} \ . \nonumber
\end{eqnarray}
For vertices involving kaons $f$ is taken to be the kaon decay constant
$f\approx 114.4$MeV. The couplings $F$ and $D$ are constrained by $F+D=g_A$; 
for the calculation we use $F=3/4$ and $D=1/2$. 

Couplings to the electromagnetic field are introduced by
\begin{eqnarray}
{\cal L}_{EM} &=& 
- e {\rm Tr}\Bigl({\overline B}[{\cal Q} A\!\!\!\!/,B]\Bigr)
- e{\mu_D\over 4M}{\rm Tr}
\Bigl({\overline B}\sigma_{\mu\nu}F^{\mu\nu}\{{\cal Q}, B\}\Bigr)
- e{\mu_F\over 4M} {\rm Tr}
\Bigl({\overline B}\sigma_{\mu\nu}F^{\mu\nu}[{\cal Q}, B]\Bigr)
\label{eq:lem}\\
&&\null 
+ e{\beta_D\over 2M^2} {\rm Tr}\Bigl({\overline B}\gamma^\nu\partial^\mu
F_{\mu\nu} \{{\cal Q} , B\}\Bigr)
+ e{\beta_F\over 2M^2} {\rm Tr}\Bigl({\overline B}\gamma^\nu\partial^\mu
F_{\mu\nu} [{\cal Q} , B]\Bigr) \ ,
\nonumber
\end{eqnarray}
where ${\cal Q}={\rm diag}\{2/3,-1/3,-1/3\}$ is the quark charge matrix.
Combined with vector meson dominance, the Lagrangian Eq.~(\ref{eq:lem}) 
describes the low energy electromagnetic structure of the baryons
so that no external form factors are needed.

To constrain the couplings we follow closely Ref.~\cite{MUELLER99}.
For the meson-baryon couplings we assume $SU(3)$ symmetry and that the OZI rule
holds, {\em i.e.} the couplings between nucleons and the $\phi$ meson
vanish. 
Furthermore, relation Eq.~(\ref{eq:mesonmix}) is implemented with
the ideal mixing angle $\sin(\theta)=1/\sqrt{3}$.
For given values of the corresponding $\omega$ and $\rho$ 
couplings to the nucleons, the set of parameters 
$(g_F, g_D, g_S, f_F, f_D, f_S)$ is then fixed ($M$ is taken to be the nucleon
mass).

The electromagnetic structure generated by Eq.~(\ref{eq:lem})
holds only in the strict $SU(3)$ limit.
The physical values of the magnetic moments can be generated 
by adding appropriate symmetry breaking terms.
We use the Particle Data Group \cite{PDG} values
\begin{equation}
\mu_{\Lambda}=-0.613  \ .
\nonumber
\end{equation}
For the magnetic moment for the $\Sigma^0$, 
which is experimentally not accessible, we employ the result
of the chiral perturbation theory calculation in Ref.~\cite{MEISSNER97}
\begin{equation}
\mu_{\Sigma^0}=0.65  \ .
\nonumber
\end{equation}
The parameters $\beta_F$ and $\beta_D$ contribute to the charge radii of 
the baryons which are not known in the hyperon sector.
For simplicity, we assume $SU(3)$ symmetry 
which determines the parameters from the corresponding values in the 
nucleon sector.
As discussed later we adjust the coupling of the $\Lambda$ to the 
scalar field $\varphi$ to obtain a good reproduction of the experimentally
known levels in $^{16}_{\Lambda}$O \cite{BRUECKNER76}.
For the $\Sigma^0$ we follow the 
phenomenological approach of Refs.~\cite{MUELLER99,SCHAFFNER94} and
require that the coupling reproduces the hyperon potential 
in nuclear matter which is taken to be
\begin{equation}
U^{\Sigma}= g^\varphi_\Sigma \varphi - g^\omega_\Sigma \omega^0 
\approx 25 {\rm MeV} \ .
\nonumber
\end{equation}
In the nucleon sector we employ the parameter set G1 of Ref.~\cite{FST97}.
The corresponding hyperon couplings are listed in Table~\ref{tab:coup}.

In principle, the Lagrangian Eq.~(\ref{eq:l0}) contains vertices that
generate a non-diagonal self energy in the $\Lambda-\Sigma^0$ sector
of flavor space. However, flavor mixing has a very small effect on the
hypernuclear energy levels \cite{MUELLER99b} and will be disregarded 
in the following.

To generate the ground state of the initial nucleus the meson field
operators in the Lagrangian Eq.~(\ref{eq:l0})
are replaced by their mean field values. 
The lowest order response functions
are then calculated in a perturbative expansion with respect
to this ground state.
%
% This is file chap3.tex
%%%%%%%%%%%%%%%%%%%%%%%%%%%
%
\section{Strangeness-changing response functions}
\label{response}
Among the many elementary processes that may produce a hyperon on a
nuclear target our theoretical analysis is guided by the hadronic and
photoproduction reactions discussed in Sec.~\ref{intro}. Yet rather
than focus on the calculation of the cross section, we consider the
nuclear response functions themselves, as they contain all essential
information on hypernuclear spectra. The linear response of the nuclear
ground state to any external probe is related to a suitably defined
polarization tensor. The polarization tensor is a fundamental
many-body operator that may be computed systematically using
well-known many-body techniques, such as Feynman diagrams and Dyson's
equation~\cite{FETWAL71}.

\subsection{Lowest-order Polarization}
\label{lowest}
To illustrate the many-body techniques employed here we simplify
the model presented in Sect.~\ref{lagrangian} and assume that there
are no meson-meson self interactions and that there is only one type
of hyperons interacting with nucleons and mesons.  Without loss of
generality we concentrate on the pseudoscalar polarization in the
following. This many-body operator is defined as the ground-state
expectation value of a time-ordered product of pseudoscalar currents
\begin{equation}
  i\Pi^{5,5}(x,y) =  \langle \Psi_{0}| 
                     T \Big[ J^{5}(x) J^{5}(y)\Big] 
                     |\Psi_{0}\rangle \;,
 \label{pi55}
\end{equation}
where 
\begin{equation}
  J^{5}(x) = \bar{\psi}_{Y}(x)i\gamma^{5}\psi_{N}(x) \;.
 \label{j5}
\end{equation}
In a mean-field approximation to the ground state the polarization
insertion can be written exclusively in terms of hyperon and nucleon
mean-field propagators
\begin{equation}
  i\Pi^{5,5}(x,y) = {\rm Tr} 
    \left[ (i\gamma^{5})G_{Y}(x,y)
           (i\gamma^{5})G_{N}(y,x) \right] \;.
 \label{pi55mf}
\end{equation}
These propagators contain information about the interaction of 
the propagating baryon with the average mean field provided by 
the nuclear medium. Note that even in a simplified description
in which the interactions are ignored, such as in a Fermi-gas 
treatment, the nucleon --- but not the hyperon --- propagator 
would still differ from its free-space value because of the 
filled Fermi sea. This fact suggests the following decomposition 
of the nucleon propagator~\cite{SERWAL86}:
\begin{eqnarray}
  &&
  G_{N}(x,y) = \int_{-\infty}^{\infty} {d\omega \over 2\pi}
               e^{-i\omega(x^{0}-y^{0})}
               G_{N}({\bf x},{\bf y};\omega) \;, \\
  &&
    G_{N}({\bf x},{\bf y};\omega)     =
    G_{F}({\bf x},{\bf y};\omega) +
    G_{D}({\bf x},{\bf y};\omega) \;.
 \label{gxy}
\end{eqnarray}
The Feynman part of the propagator, $G_{F}$, admits a spectral 
decomposition in terms of the mean-field solutions to the Dirac 
equation. That is, 
\begin{equation}
  G_{F}({\bf x},{\bf y};\omega) = \sum_{\alpha}
   \left[
     \frac{U_{\alpha}({\bf x})\overline{U}_{\alpha}({\bf y})}
          {\omega - E_{\alpha}^{(+)} + i\eta} + 
     \frac{V_{\alpha}({\bf x})\overline{V}_{\alpha}({\bf y})}
          {\omega + E_{\alpha}^{(-)} - i\eta}  
   \right] \;,
 \label{gfeyn}
\end{equation}
where $U_{\alpha}$ and $V_{\alpha}$ are the positive- 
and negative-energy solutions to the Dirac equation, and 
the sum is over all states in the spectrum. The density-dependent
part of the propagator, $G_{D}$, corrects $G_{F}$ for the presence 
of a filled Fermi surface. Formally, one effects this correction 
by shifting the position of the pole of every occupied state from 
below to above the real axis
\begin{eqnarray}
    G_{D}({\bf x},{\bf y};\omega) &=& \sum_{\alpha < {\rm F}}
     U_{\alpha}({\bf x})\overline{U}_{\alpha}({\bf y}) 
     \left[
        {1 \over \omega - E_{\alpha}^{(+)} - i\eta} -
        {1 \over \omega - E_{\alpha}^{(+)} + i\eta} 
     \right] \\ &=&
     2 \pi i \sum_{\alpha < {\rm F}}
     \delta\Big(\omega-E_{\alpha}^{(+)}\Big)     
               {U}_{\alpha}({\bf x})        
      \overline{U}_{\alpha}({\bf y})        \;.
 \label{gdens}
\end{eqnarray}
Note that the sum over $\alpha$ is now restricted to only those
positive-energy nucleon states below the Fermi surface. The hyperon 
propagator, because of the $S\!=\!0$ nature of the nuclear ground 
state, does not suffer from such a correction.

The decomposition of the nucleon propagator into Feynman and
density-dependent contributions suggests an equivalent decomposition
for the polarization insertion
\begin{eqnarray}
  &&
  \Pi^{5,5}(x,y) = \int_{-\infty}^{\infty} {d\omega \over 2\pi}
              e^{-i\omega(x^{0}-y^{0})}
             \Pi^{5,5}({\bf x},{\bf y};\omega) \;, \\
  &&
  \Pi^{5,5}({\bf x},{\bf y};\omega)     =
  \Pi^{5,5}_{F}({\bf x},{\bf y};\omega) +
  \Pi^{5,5}_{D}({\bf x},{\bf y};\omega) \;.
 \label{pixyfd}
\end{eqnarray}
The Feynman part of the polarization, $\Pi^{5,5}_{F}$, describes pair
production --- or vacuum polarization. In infinite nuclear matter, the
threshold for pair production lies well into the timelike region at
$q^{2}=(M_{Y}^{*}+M_{N}^{*})^{2}$ ($M^{*}$ is the effective baryon
mass in the nuclear medium). This is far away from the spacelike
region accessible in hadronic and photoproduction processes. Thus the
lowest-order response --- which is proportional to the imaginary part
of the polarization insertion --- is not sensitive to pair production
and will be ignored henceforth. Yet this statement should be taken
with caution as we are interested in computing the nuclear response 
beyond first order. Although the imaginary part of vacuum polarization 
does indeed vanish for spacelike processes, its dispersive (real) 
content does not: vacuum excitations can be produced virtually. Thus 
in more sophisticated treatments of the response than the one
presented here, pair production may play a significant role. Still,
the RPA response presented here is fully consistent with the
``no-vacuum'' approximation to the mean-field ground state.

In contrast to the Feynman part of the polarization, which is divergent
and must be renormalized, the density-dependent part is finite and 
given by
\begin{equation}
  \label{pidall}
  \Pi_{D}^{5,5}({\bf x},{\bf y};\omega) =
   \sum_{\alpha<F} 
   \overline{U}_{\alpha}({\bf x})(i\gamma^{5})
    G_{Y}\Big({\bf x},{\bf y};E_{\alpha}^{(+)}+\omega\Big) 
    (i\gamma^{5})U_{\alpha}({\bf y})  \;.
   \label{pidens}
\end{equation}
The density-dependent part of the polarization describes the 
traditional excitation of particle-hole pairs. This may be
most clearly seen by using a spectral decomposition of the 
hyperon propagator as in Eq.~(\ref{gfeyn})
\begin{equation}
  \Pi_{D}^{5,5}({\bf x},{\bf y};\omega) =
   \sum_{\alpha<F,\beta} 
   \left[
     {\overline{U}_{\alpha}({\bf x})(i\gamma^{5}) 
     {\cal U}_{\beta}({\bf x})\overline{\cal U}_{\beta}({\bf y})
     (i\gamma^{5})U_{\alpha}({\bf y}) \over
     \omega - \Big({\cal E}^{+}_{\beta}-E^{+}_{\alpha}\Big) + i\eta} +
     {\overline{U}_{\alpha}({\bf x})(i\gamma^{5})
     {\cal V}_{\beta}({\bf x})\overline{\cal V}_{\beta}({\bf y})
     (i\gamma^{5})U_{\alpha}({\bf y}) \over
    \omega + \Big({\cal E}^{-}_{\beta}+E^{+}_{\alpha}\Big) - i\eta} 
   \right] \;.
 \label{pidspect}
\end{equation}   
The first term in the sum represents the formation of a
$Y$-particle--$N$-hole pair after the probe has transfered an energy
$\omega$ to the nucleus. The excitation of the pair becomes real, 
namely both particles get on their mass shell, only when the energy 
transfer is identical to the excitation energy 
$\omega\equiv{\cal E}^{+}_{\beta}-E^{+}_{\alpha}$. That is, the 
inclusive pseudoscalar response to lowest order becomes
\begin{equation}
  S_{\rm ps}({\bf q},\omega) = - \frac{1}{\pi}{\rm Im}
    \left[\Pi_{D}^{5,5}({\bf q},{\bf q};\omega)\right]
    =\sum_{\alpha<F,\beta}
    \left|\rho^{(5)}_{\beta\alpha}(q)\right|^{2}
    \delta(\omega-\omega_{\beta\alpha}) \;.
 \label{impi}
\end{equation}
Note that we have introduced the transition density
$\rho^{(5)}_{\beta\alpha}$ and the excitation energy 
$\omega_{\beta\alpha}$ by
\begin{equation}
  \rho^{(5)}_{\beta\alpha}(q)\equiv
    \int d^3x \,e^{i{\bf q}\cdot{\bf x}}\,
    \overline{\cal U}_{\beta}({\bf x})
    (i\gamma^{5})U_{\alpha}({\bf x}) \quad \mbox{and} \quad
    \omega_{\beta\alpha}\equiv
    {\cal E}^{+}_{\beta}-E^{+}_{\alpha} \;.
 \label{rho5def}
\end{equation}
The second term in the sum is interesting and has no nonrelativistic
counterpart; it represents the Pauli blocking of the vacuum
excitations discussed earlier. This term mixes positive (nucleon) and
negative (hyperon) states and prevents the nucleon --- from the
$N\bar{Y}$ pair --- from occupying a state below the Fermi surface. As
in the case of vacuum polarization, this term is purely real and makes
no contribution to the lowest-order response. Yet its inclusion in the
correlated RPA response is essential in order to satisfy fundamental 
symmetries of nature, such as gauge invariance.
\subsection{The RPA equations}
\label{rpa}
In our simplified version of the model, 
the polarization tensor also describes
modifications to the propagation of mesons (such as the $K^{-}$ meson),
in addition to containing all information about the excitation
spectra of hypernuclei.
In the lowest-order
approximation, the residual interaction between the particle and the
hole is neglected. A consistent approximation scheme that goes beyond
lowest order --- by including the residual interaction between the
nucleon-hole and the hyperon-particle is the random-phase
approximation. In the RPA one includes many-body correlations by
iterating to infinite order the lowest-order polarization
insertion. 
For example, assuming a pseudo-scalar coupling, this means
\begin{equation}
  \Pi^{5,5}_{\rm RPA}({\bf q},{\bf q}';\omega) =
  \Pi^{5,5}({\bf q},{\bf q}';\omega) +
  \int\frac{d^3k}{(2\pi)^{3}}
  \Pi^{5,5}({\bf q},{\bf k};\omega)           
  V_{K}({\bf k};\omega)
  \Pi^{5,5}_{\rm RPA}({\bf k},{\bf q}';\omega) \;,
 \label{Pi55RPA}
\end{equation}
where we have introduced the free kaon propagator through
\begin{equation}
  V_{K}({\bf k};\omega) = 
  -g_{YNK}^{2}\Delta_{K}({\bf k},\omega)=
  \frac{-g_{YNK}^{2}}{\omega^{2}-{\bf k}^{2}-m_{K}^{2}} \;.
 \label{Kpropagator}
\end{equation}
Note that $g_{YNK}$ represents the hyperon-nucleon-kaon coupling
constant.  That the RPA is the consistent linear response of the
mean-field ground state has been amply demonstrated through the
excitation of the isoscalar-dipole ($J^{\pi}\!=\!1^{-};T\!=\!0$)
mode. The isoscalar-dipole response provides a delicate test of the
self-consistency of the calculation. Indeed, a calculation of the
isoscalar-dipole response to lowest-order is flawed by the presence 
of spurious strength associated with the uniform translation of the
center of mass. In contrast, a consistent RPA calculation --- one
that uses the same interaction in the calculation of the mean-field
ground-state than in the calculation of the response --- eliminates
this anomalous behavior by shifting all spurious strength to zero
excitation energy.

In the previous section a spectral decomposition of the Feynman
propagator was presented. Although such a decomposition is useful for
understanding the spectral content of the nuclear response, in
practice it suffers from too much reliance on artificial cutoffs. An
efficient scheme that avoids the introduction of artificial cutoffs is
the non-spectral approach. Moreover, a non-spectral formalism has the
added advantage that, at least in principle, the positive-energy
continuum may be treated exactly. The non-spectral generation of the
Feynman propagator amounts to solving the following inhomogeneous
Dirac equation
\begin{equation}
  \left[ 
   \omega\gamma^{0}+i{\bf\gamma}\cdot{\bf \nabla}
    -M_B-\Sigma({\bf x})
  \right]G_{F}({\bf x},{\bf y};\omega)=
  \delta({\bf x}-{\bf y}) \;,
 \label{DiracEq}
\end{equation}
where $\omega$ is in general a complex variable and the mean-field 
potential, or baryon self-energy, is given by
\begin{equation}
 \Sigma({\bf x}) = \Sigma_{s}({\bf x}) 
                 + \gamma^{0}\Sigma_{0}({\bf x})
                 -\frac{i}{2M}
                  \gamma^{0}\frac{{\boldmath\gamma}\cdot{\bf x}}{|{\bf x}|}
                  \Sigma_{T}({\bf x}) \;.
\label{Sigma}
\end{equation}
The quantities $\Sigma_{s}$, $\Sigma_{0}$ 
and $\Sigma_{T}$ contain the nuclear 
scalar, vector and tensor potentials generated at the mean-field level.
These mean-field self-energies have been collected in 
Table~\ref{tab:pot}. 
Note that we have assumed that the mean-field 
potential is generated by a spherically-symmetric, spin-saturated 
ground state. Taking advantage of this spherical symmetry one may 
decompose the Feynman propagator in terms of spin-spherical 
harmonics
\begin{equation}
  G_{F}({\bf x},{\bf y};\omega)=\frac{1}{xy}\sum_{\kappa m}
   \left(
   \matrix{
           \phantom{i}
           g_{11}^{\kappa}(x,y;\omega)
           \langle\hat{\bf x}|\!+\!\kappa m\rangle
           \langle +\kappa m |\hat{\bf y}\rangle &
         -ig_{12}^{\kappa}(x,y;\omega)
           \langle\hat{\bf x}|\!+\!\kappa m\rangle
           \langle -\kappa m |\hat{\bf y}\rangle \cr
          ig_{21}^{\kappa}(x,y;\omega)
           \langle\hat{\bf x}|\!-\!\kappa m\rangle
           \langle +\kappa m |\hat{\bf y}\rangle &
           \phantom{-i}
           g_{22}^{\kappa}(x,y;\omega)
           \langle\hat{\bf x}|\!-\!\kappa m\rangle
           \langle -\kappa m |\hat{\bf y}\rangle }
   \right) \;,
 \label{gij}
\end{equation}
which are defined as
\begin{mathletters}
\begin{eqnarray}
  &&
  \langle\hat{\bf x}|\kappa m\rangle  =
  \sum_{m_{l}m_{s}}
  \langle lm_{l},{\scriptstyle{\frac{1}{2}}}m_{s}|
  l{\scriptstyle{\frac{1}{2}}}jm\rangle 
  Y_{lm_{l}}(\hat{\bf x})\chi_{{\scriptstyle{\frac{1}{2}}}m_{s}}\;,
 \label{curlyy} \\
  && \,
  j\!=\!|\kappa|\!-\!\frac{1}{2}\; \quad\hbox{and}\quad
  l=\cases{+\kappa   & if $\kappa>0\;,$ \cr
           -\kappa-1 & if $\kappa<0\;.$   }
 \label{ljkappa}
\end{eqnarray}
\end{mathletters}
The above decomposition enables one to rewrite the Dirac equation
as a set of first-order, coupled, ordinary differential equations 
of the form
\begin{equation}
  \left(
   \matrix{
   \omega^{*}\!-\!M^{*} & 
   \displaystyle{\frac{d}{dx}\!-\!\frac{\kappa^{*}}{x}} \cr
   \displaystyle{\frac{d}{dx}\!+\!\frac{\kappa^{*}}{x}} &
  -\omega^{*}\!-\!M^{*} }
  \right)
  \left(
   \matrix{g_{11}^{\kappa} & g_{12}^{\kappa} \cr
                           &                 \cr
           g_{21}^{\kappa} & g_{22}^{\kappa} }
  \right)=\delta(x-y) \;,
 \label{DiracEqG}
\end{equation}
where we have defined
\begin{equation}
  \omega^{*}\equiv \omega-\Sigma_{v}(x) \;,\;
  M^{*}\equiv M_B+\Sigma_{s}(x)\;, \;
  \;\hbox{and}\;\;\;
  \kappa^{*}\equiv \kappa-\frac{x}{2M}
  \Sigma_{T}(x)\;.
 \label{stars}
\end{equation}
Similarly, a positive-energy Dirac spinor 
\begin{equation}
  U_{\alpha}({\bf x})=\frac{1}{x}
   \left(
   \matrix{
           \phantom{i}g_{n\kappa}(x)
           \langle\hat{\bf x}|\!+\!\kappa m\rangle \cr
           if_{n\kappa}(x)
           \langle\hat{\bf x}|\!-\!\kappa m\rangle }
   \right)  \quad (\alpha\equiv{n\kappa m}) \;,
 \label{uspinor}
\end{equation}
satisfies the homogeneous Dirac equation 
\begin{mathletters}
 \begin{eqnarray}
  &&\left(\frac{d}{dx}+\frac{\kappa^{*}}{x}\right)g_{n\kappa}(x)
    -\left(E^{*}+M^{*}\right)f_{n\kappa}(x)=0 \;, \\
  &&\left(\frac{d}{dx}-\frac{\kappa^{*}}{x}\right)f_{n\kappa}(x)
    +\left(E^{*}-M^{*}\right)g_{n\kappa}(x)=0 \;.
 \end{eqnarray}
 \label{DiracEqU}
\end{mathletters}

At the beginning of this section we expressed Dyson's equation 
for the pseudoscalar polarization as an integral equation in 
three dimensions [see Eq.~(\ref{Pi55RPA})]. Yet the spherical 
symmetry of the problem simplifies this numerical task considerably. 
To do so we used the results derived earlier for the Feynman
propagator and for the positive-energy Dirac spinors to
perform a multipole decomposition of the polarization 
insertion as described in Ref.~\cite{HOROWITZ90}:
\begin{equation}
  \Pi^{5,5}({\bf q},{\bf q}';\omega) = \sum_{J=0}^{\infty}
  \Pi^{5,5}_{J}(q,q';\omega)
   P^J_{00}({\hat{\bf q}},{\hat{\bf q}'}) \;.
 \label{multipole}
\end{equation}
Here the angular dependence is fully contained in the function,
\begin{equation}
   P^J_{\lambda\lambda^\prime}({\hat{\bf q}},{\hat{\bf q}'})=
   \sum_{M} D^{J }_{M\lambda}({\hat{\bf q}})
            D^{J*}_{M\lambda^\prime}({\hat{\bf q}'})\;, 
 \label{pfunction}
\end{equation}
which is expressed in terms of Wigner D-functions~\cite{EDMONDS57}.
The multipoles of the polarization insertion, $\Pi^{5,5}_{J}$, 
involve various reduced matrix elements enforcing selection rules 
for the intermediate hyperon-particle and nucleon-hole states that 
couple to a total angular momentum $J$. Note that because of the 
pseudoscalar nature ($\gamma_5$) of the vertex, unnatural parity 
states $(J^\pi=0^-,1^+,\ldots)$ are created exclusively. One can
easily show, as it has been done Ref.~\cite{HOROWITZ90}, that by 
using the following identity
\begin{equation}
   \int d{\hat{\bf k}}
   P^J_{\lambda\sigma}({\hat{\bf q}},{\hat{\bf k}})
   P^{J^\prime}_{\sigma\lambda^\prime}({\hat{\bf k}},{\hat{\bf q}'}) =
   {4\pi\over 2J+1} \delta_{JJ^\prime}
   P^{J}_{\lambda\lambda^\prime}({\hat{\bf q}},{\hat{\bf q}'}) \;,
\label{orthop}
\end{equation}
the three-dimensional integral equation for the RPA polarization 
[Eq.~(\ref{Pi55RPA})] can now be reduced to a one-dimensional one, 
albeit one for each value of $J$:
\begin{equation}
  \Pi^{5,5}_{{\rm RPA},J}(q,q';\omega) =
  \Pi^{5,5}_{J}(q,q';\omega) +
  \frac{1}{2J\!+\!1}
  \int\frac{dk}{2\pi^2}
  \Pi^{5,5}_{J}(q,k;\omega)           
   V_{K}(k;\omega)
  \Pi^{5,5}_{{\rm RPA},J}(k,q';\omega) \;.
 \label{PiJ55RPA}
\end{equation}
\subsection{Residual particle-hole interaction}
\label{residual}
The RPA equation (\ref{Pi55RPA}) is only applicable in the simplified
version of the hadronic model where the full response function Eq.(\ref{pi55})
is equivalent to the full kaon propagator.
In general, the connection between response functions
and full meson propagators is more involved because of
kaon and pion self-interactions and because of the fact that 
different hyperon flavors can couple to the same external source.
Taking into account the possibility of creating a particle-hole state
with a $\Lambda$ or a $\Sigma^0$ hyperon we consider response functions of 
the type
\begin{equation}
\Pi^{\alpha,\beta}(x,y)=\sum_{Y=\Lambda,\Sigma^0}
\langle\Psi_0|T[{\overline\Psi}_N(x)\Gamma^\alpha_Y\Psi_Y(x)
{\overline \Psi}_Y(y)
\Gamma^\beta_Y\Psi_N(y)]|\Psi_0\rangle
\label{responseab}
\end{equation}
where $\Gamma^\alpha_Y$ characterizes specific interaction vertices.
Although these response functions can be analyzed for arbitrary
vertices in terms of Feynman diagrams it is more intuitive to cast the
problem into a form similar to Eq.~(\ref{Pi55RPA}).  In analogy to
Eq.~(\ref{pixyfd}) we keep only the density-dependent part for the
lowest order contribution to the response function (\ref{responseab}).
As a further approximation we neglect internal vertices that give rise
to meson-meson self interactions and loops involving $\Sigma^\pm$ and
$\Xi^{\overline{0}}$ baryons.  These contributions constitute pure
``vacuum'' contributions as discussed in Sec.~\ref{lowest}.  The
remaining piece to be specified is then the residual hyperon-nucleon
interaction.  Although considerable progress has been made towards
elucidating the precise form of the interaction, much work remains to
be done.  Thus in this, our first contribution to the subject, we make
a very simple assumption as to the nature of the hyperon-nucleon
residual interaction. We assume, in analogy with the ``$\pi$+$\rho$''
isovector interaction, that the residual interaction in the
hyperon-nucleon channel is mediated by the two lightest $S\!=\!-1$
mesons: the pseudoscalar kaon and its vector partner the $K^{*}(892)$.
In this case the range of the interaction is determined from the free
propagators. These are given by
\begin{mathletters}
 \begin{eqnarray}
  \Delta_{K}(q) &=& \frac{1}{q^{2}-m_{K}^{2}} \;, \\
   D^{\mu\nu}_{K^{\!*}}(q) &=& 
   \frac{-g^{\mu\nu}+q^{\mu}q^{\nu}/m_{K^{*}}^2}
        {q^{2}-m_{K^{*}}^{2}} \;.
 \label{ksfree}
 \end{eqnarray}
\end{mathletters}
The strength and the spin structure of the residual interaction is,  
on the other hand, determined by the vertices. For the $YNK$ vertex 
we assume a pseudovector --- as opposed to a pseudoscalar --- 
representation, as detailed in Eq.~(\ref{eq:lmb}).  Although both 
representations are equivalent on-shell, it is convenient to adopt 
the pseudovector representation as it accounts for the correct 
low-energy theorems --- especially in the case of the pion, the
lightest member of the pseudoscalar octet --- without the need for 
sensitive cancellations. The $K^{*}$-meson contains a vector as 
well as a tensor coupling to the baryons. These choices specify 
completely the nature of the elementary $YNK$ and $YNK^{*}$ 
vertices:
\begin{equation}
  \Gamma^{\alpha}_Y = 
  \cases{ 
   \displaystyle{\frac{g^{K}_{YN}}{f_K}{\rlap/{q}}\gamma^{5}} 
    & if $\alpha=K$; \cr
   \displaystyle{g^{K^{\!*}}_{YN}\gamma^{\mu}
               +if^{K^{\!*}}_{YN}\,\sigma_{\mu\nu}
               \frac{q_{\nu}}{2M}}
    & if $\alpha=K^{*}$. \cr} 
 \label{vertex}
\end{equation}
Choosing $\alpha=\beta=K$ and $\alpha=\beta=K^{*}$ in Eq.~(\ref{responseab})
defines a $K$ and a $K^*$ response function, respectively.
With the approximation outlined above, each response functions can 
then be obtained from a RPA equations similar to Eq.~(\ref{Pi55RPA}).

The spin structure of the vertices is well known, 
at least in the nonrelativistic limit. Indeed, as in the case of the pion, 
the kaon generates a spin-longitudinal coupling of the form
\begin{equation}
 \displaystyle{\frac{g^{K}_{YN}}{f_K}{\rlap/{q}}\gamma^{5}} 
 \rightarrow \frac{g^{K}_{YN}}{f_K}({\bf \sigma}\cdot{\bf q}) \;.
 \label{spinlong}
\end{equation}
Thus, the kaon excites exclusively hypernuclear states
of unnatural parity. In contrast, the vector $K^{*}$-meson
induces, in addition to the simple spin-independent coupling 
stemming from the timelike part of Dirac vertex, a spin-transverse 
coupling of the form
\begin{mathletters}
 \begin{eqnarray}
   &&
   g^{K^*}_{YN}{\boldmath\gamma} \rightarrow
   \frac{g^{K^*}_{YN}}{2M}({\bf \sigma}\times{\bf q}) \;, \\
   &&
   f^{K^*}_{YN}\sigma_{\mu\nu}\frac{q_{\nu}}{2M}  
   \rightarrow
   \frac{f^{K^*}_{YN}}{2M}({\bf \sigma}\times{\bf q}) \;.
 \label{spintrans}
 \end{eqnarray}
\end{mathletters}
Hence, with the exception of $0^{-}$ states, a $K^{*}$-meson
can excite hypernuclear states of all spins and parities. Note 
that because of the present lack of theoretical guidance, no
repulsive short-range Landau-Migdal parameter has been introduced. 
Thus, in this simplified description there is no source of mixing 
between the spin-longitudinal and the spin-transverse modes.
%
%
% This is file chap4.tex
%%%%%%%%%%%%%%%%%%%%%%%%%%%
%
\section{Results and discussion}
\label{results}
Let us start the discussion with the $K^*$ response and assume that
the hypernuclear states are created by replacing a neutron in the
initial nucleus with a hyperon.  Although the calculated response
function contains information on both $\Lambda$ and $\Sigma^0$ states,
we will restrict the discussion to energies well below the threshold
for creating $\Sigma^0$ hypernuclei.

In analogy to the electromagnetic case, the $K^*$ response
can be analyzed in terms of structure functions. Unlike the electromagnetic 
response, however, there are generally four structure functions because the
$K^*$ does not couple to a conserved current. Here we focus on the longitudinal
and transverse response which are defined by
\begin{eqnarray}
\Pi_L(q_0,{\bf q},{\bf q})&=&\Pi^{00}(q_0,{\bf q},{\bf q}) \ , \label{long} \\
\Pi_T(q_0,{\bf q},{\bf q})&=&\Pi_i^i(q_0,{\bf q},{\bf q})
-\frac{q_iq_j}{|\bf q|}\Pi^{ij}(q_0,{\bf q},{\bf q}) \ ,
\label{trans}
\end{eqnarray}
and that contain all the relevant information on the spectra.

\begin{figure}[tbhp]
\centerline{\vbox to 0.5in{}}
\centerline{\psfig{figure=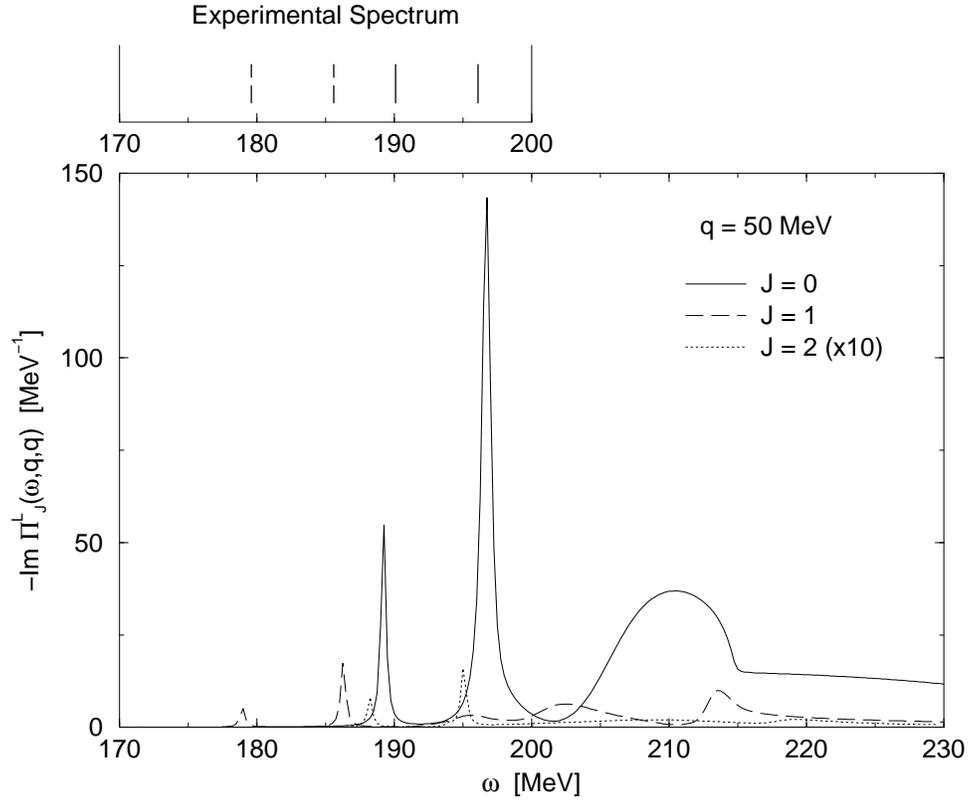,width=5in}}
\centerline{\vbox to 0.2in{}}
\caption{Lowest order multipoles of the longitudinal $K^{*}$ response
in oxygen.}
\centerline{\vbox to 0.5in{}}
\label{fig:ox1}
\end{figure}

In hadronic reactions of the type $(K^-,\pi^-)$ 
hypernuclear states are effectively populated if
the outgoing meson is emitted in the forward direction. Consequently,
these reactions are characterized by small momentum transfer $(|{\bf
q}|< q_0)$.  The RPA results for the longitudinal $K^{*}$ response in
oxygen are indicated in Fig.\ref{fig:ox1}.  At low momentum transfer
($|{\bf q}|= 50$MeV) the multipole expansion is rapidly converging and
only the lowest order terms are indicated.  The coupling of the scalar
$\varphi$-meson was adjusted to reproduce the two $0^+$ and $1^-$
states that are experimentally identified in the $^{16}$O
$(K^-,\pi^-)\, ^{16}_\Lambda$O reaction \cite{BRUECKNER76}.
According to Eq.~(\ref{pidspect}), we added a small imaginary part
$(i\eta=i.2$MeV) to the energy variable to keep the magnitude of the
resonances finite.  In the notation of Eq.~(\ref{impi}), the binding
energy $B_\Lambda$ of a $\Lambda$ at the location of a resonance may
be obtained from
\begin{equation}
\omega_{\Lambda0}={\cal E}^{(+)}_{\Lambda}-E^{(+)}_0\equiv B_n-B_\Lambda 
+ M_\Lambda-M_N \ ,
\label{bl}
\end{equation}
where $B_n$ denotes the separation energy of the weakest bound neutron. 

\begin{figure}[tbhp]
\centerline{\vbox to 0.5in{}}
\centerline{\psfig{figure=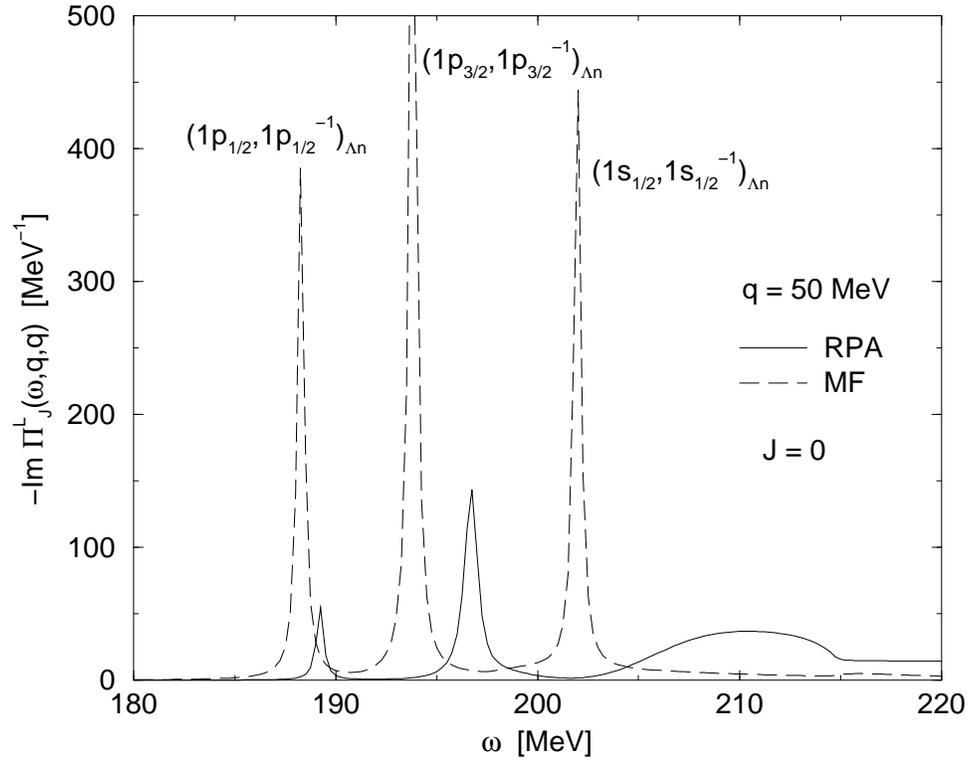,width=5in}}
\centerline{\vbox to 0.2in{}}
\caption{Mean field and RPA result for the
$J=0$ multipole of the longitudinal $K^{*}$ response in oxygen.}
\centerline{\vbox to 0.5in{}}
\label{fig:ox2}
\end{figure}

The differences between the RPA and the uncorrelated mean field predictions
can be studied in Fig.\ref{fig:ox2} for the $J=0$ multipole.
The uncorrelated states are indicated by their particle-hole contents.
The resulting RPA states are an admixture of all states with
a given total angular momentum $J$.
The increase of the hypernuclear
binding energies indicates the repulsive character of the residual interaction.
Most remarkably, there is no clear remnant of the 
$(1s_{1/2},1s_{1/2}^{-1})$ state which has moved into the continuum.
This is an important feature
because this state is predicted by the simple particle-hole
model but is not recognized in the experimental data.

\begin{figure}[tbhp]
\centerline{\vbox to 0.5in{}}
\centerline{\psfig{figure=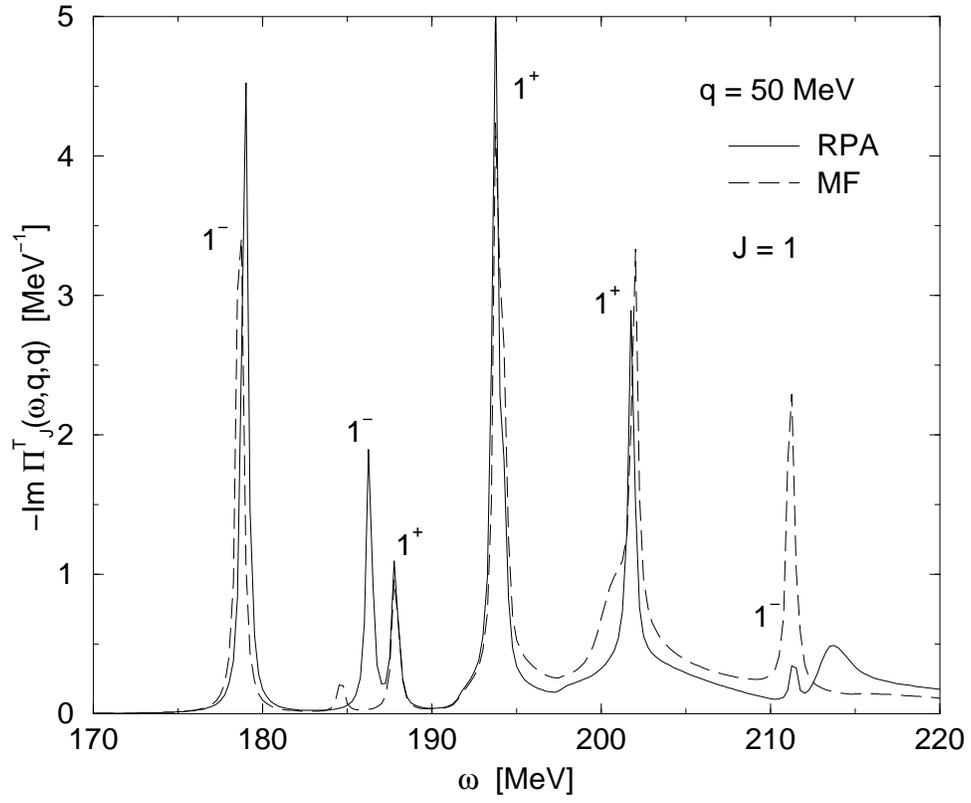,width=5in}}
\centerline{\vbox to 0.2in{}}
\caption{Mean field and RPA result for the
$J=1$ multipole of the transverse $K^{*}$ response in oxygen.}
\centerline{\vbox to 0.5in{}}
\label{fig:ox5}
\end{figure}

The longitudinal response leads exclusively to states
with natural parity. Excitations with unnatural parity arise in the
transverse response which is indicated in Fig.~\ref{fig:ox5}.
In analogy to the electromagnetic case, transverse modes with unnatural parity
can be interpreted as magnetic excitations, whereas
the modes in Fig.~\ref{fig:ox5}
with opposite parity are of transverse-electric character.
Compared to the longitudinal response, the strength of the transverse
excitations is much smaller.
Furthermore, correlation effects are modest
as a consequence of the spin-transverse coupling 
indicated in Eq.~(\ref{spintrans}).

\begin{figure}[tbhp]
\centerline{\vbox to 0.5in{}}
\centerline{\psfig{figure=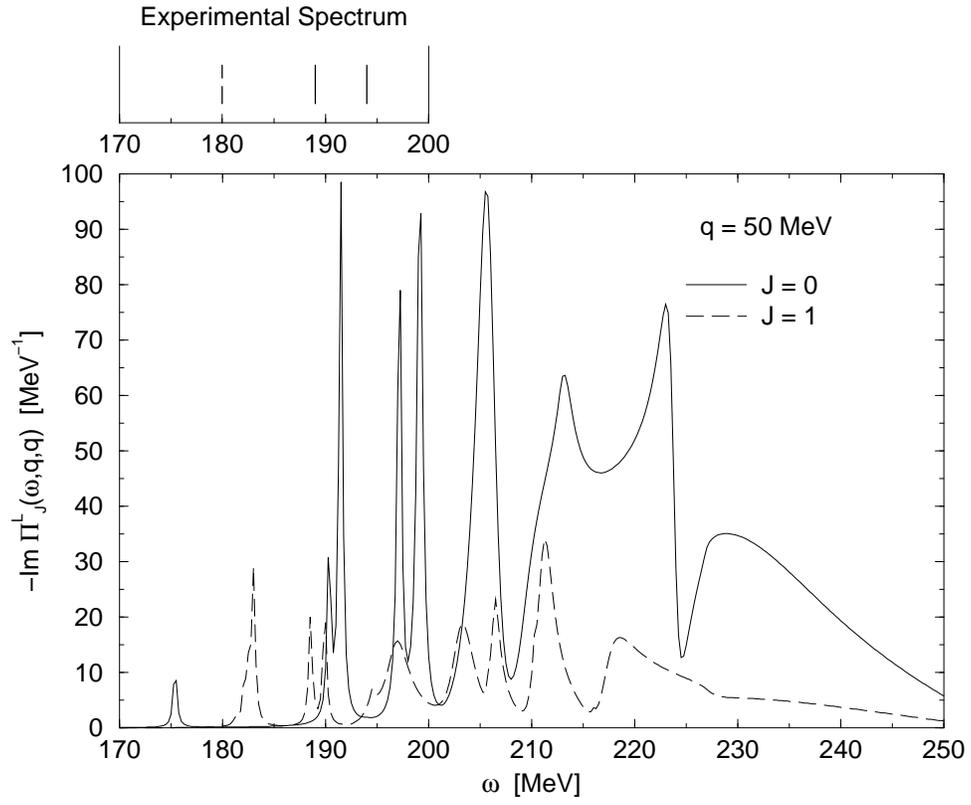,width=5in}}
\centerline{\vbox to 0.2in{}}
\caption{Lowest order multipoles of the longitudinal $K^{*}$ response in
calcium.}
\centerline{\vbox to 0.5in{}}
\label{fig:ca1}
\end{figure}

\begin{figure}[tbhp]
\centerline{\vbox to 0.5in{}}
\centerline{\psfig{figure=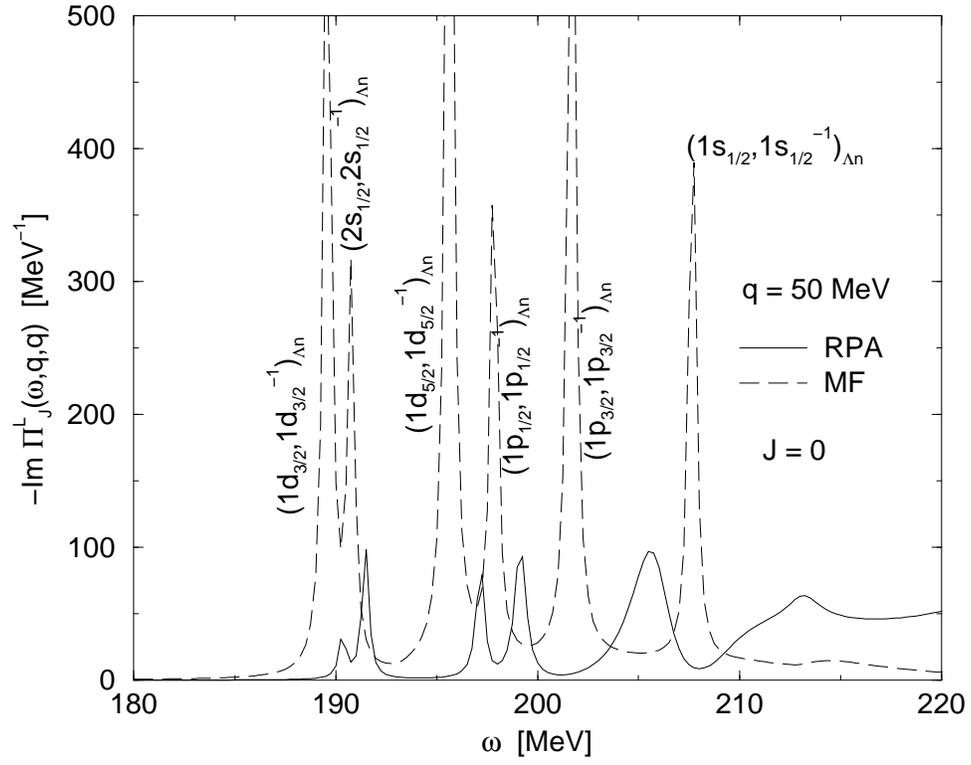,width=5in}}
\centerline{\vbox to 0.2in{}}
\caption{Mean field and RPA result for the
$J=0$ multipole of the longitudinal $K^{*}$ response in calcium.}
\centerline{\vbox to 0.5in{}}
\label{fig:ca2}
\end{figure}

The RPA predictions for calcium are shown in Fig.~\ref{fig:ca1}. 
The experimental spectrum \cite{POVH80} observed in the $(K^-,\pi^-)$ reaction
on $^{40}$Ca exhibits three peaks. Two at 
${\cal E}^{(+)}_{\Lambda}-E^{(+)}_0=194$ and 189 MeV which in the simple 
particle-hole picture are identified as
$0^+$ states with the particle-hole assignments 
$(1d_{5/2},1d_{5/2}^{-1})_{\Lambda n}$ and
$(1d_{3/2},1d_{3/2}^{-1})_{\Lambda n}$, respectively.
Furthermore, a peak at 180 MeV which is identified as a $1^-$ state with
the assignment $(1p_{1/2},1d_{3/2}^{-1})_{\Lambda n}$.
However, because of the strong mixing it is not possible to attribute the peaks
to specific states if correlations are taken into account. To make contact
with the experimental observations Fig.~\ref{fig:ca2} shows the
difference between the mean field result and the RPA prediction.
We remind the reader that the very small width of the states is an idealization
and that in a more realistic scenario the peaks are considerably 
broader. It can be expected that the two group of states around 191 MeV and
197 MeV in the RPA coalesce to form the resonance structure found in the 
experimental data. Similar as in the case of oxygen, the remnant of the
particle-hole state with the neutron in the deepest bound $s$ shell moves into
the continuum.

\begin{figure}[tbhp]
\centerline{\vbox to 0.5in{}}
\centerline{\psfig{figure=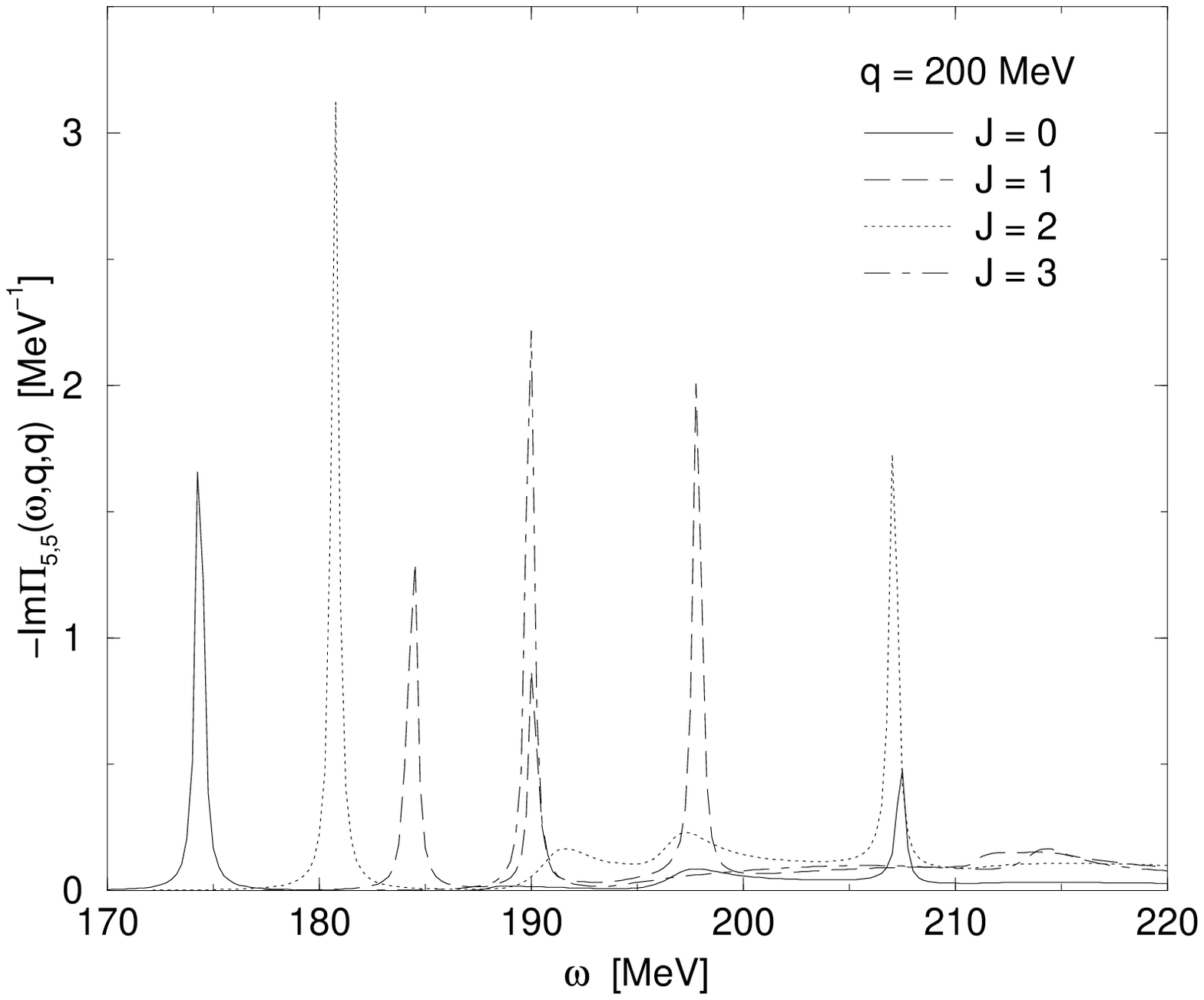,width=5in}}
\centerline{\vbox to 0.2in{}}
\caption{Lowest order multipoles of the kaon response in
$^{16}{\rm O} \rightarrow ^{16}_\Lambda{\rm N}$.}
\centerline{\vbox to 0.5in{}}
\label{fig:ox3}
\end{figure}

\begin{figure}[tbhp]
\centerline{\vbox to 0.5in{}}
\centerline{\psfig{figure=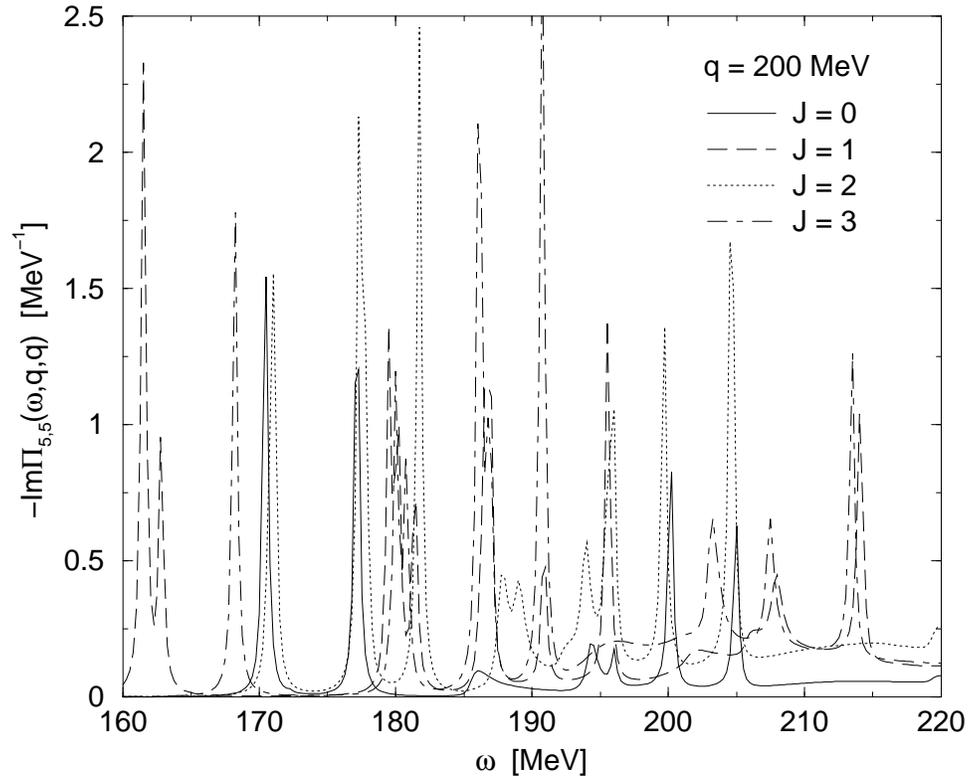,width=5in}}
\centerline{\vbox to 0.2in{}}
\caption{Lowest order multipoles of the kaon response in
$^{40}{\rm Ca} \rightarrow ^{40}_\Lambda{\rm K}$.}
\centerline{\vbox to 0.5in{}}
\label{fig:ca3}
\end{figure}

\begin{figure}[tbhp]
\centerline{\vbox to 0.5in{}}
\centerline{\psfig{figure=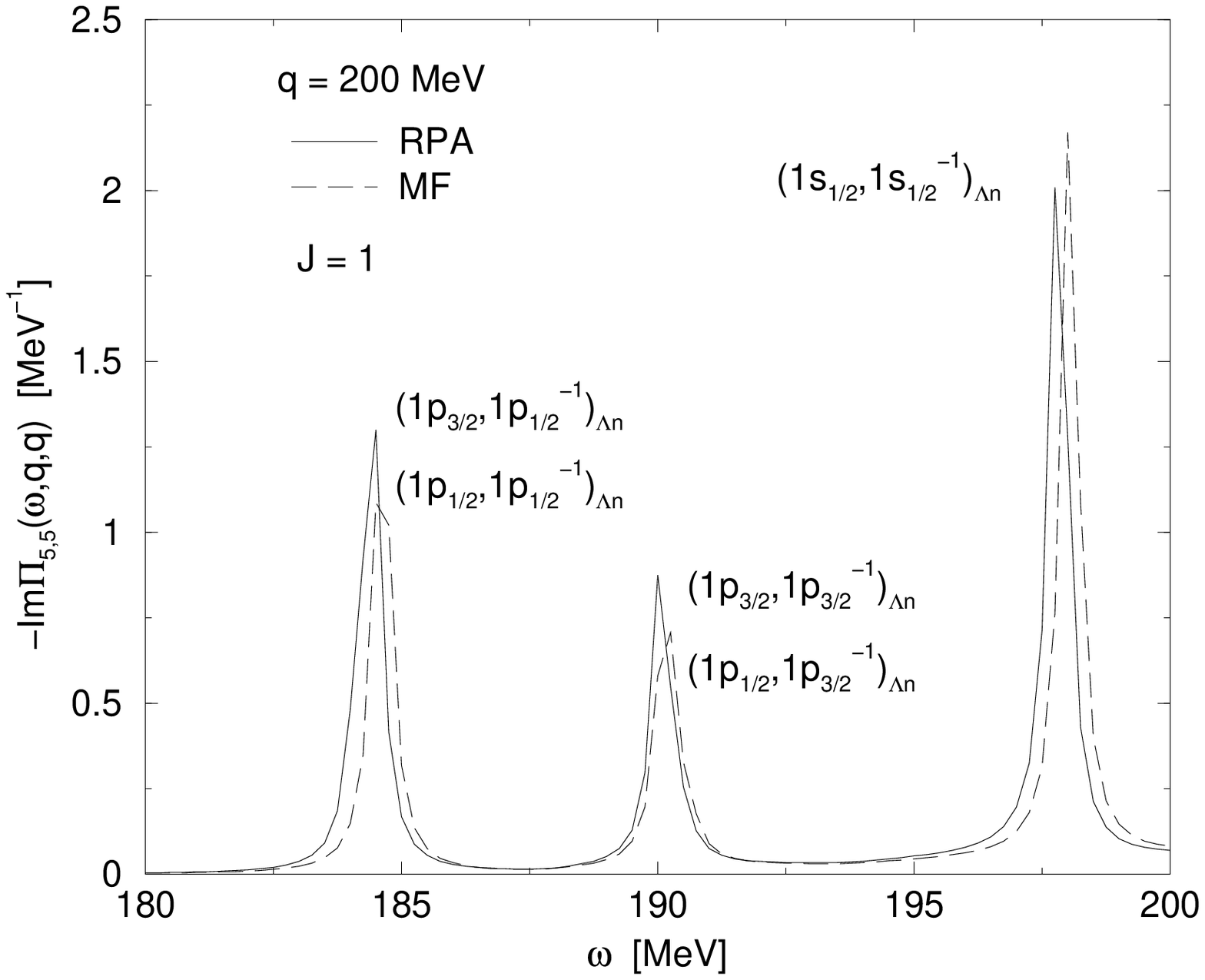,width=5in}}
\centerline{\vbox to 0.2in{}}
\caption{Mean field and RPA result for the
$J=1$ multipole of the kaon response in
$^{16}{\rm O} \rightarrow ^{16}_\Lambda{\rm N}$.}
\centerline{\vbox to 0.5in{}}
\label{fig:ox4}
\end{figure}

Based on hadronic reactions, transverse modes, including states of unnatural 
parity, are of minor importance in hypernuclear spectroscopy.
Because of their small strength, these states are usually not observed in the 
kinematic range covered by the experiments \cite{CHRIEN79}.
As an attractive alternative, states of unnatural parity 
may be studied in electromagnetic production of hypernuclei.
In our approach unnatural parity states can exclusively be studied
in the kaon response function.
Under the assumption that the
hypernuclei states are created in hyperon photoproduction off a proton,
a different range for the momentum transfer has to be considered. 
With the kinematics forced
on the problem the momenta are generally much higher than in the hadronic
reactions. However, for the outgoing kaon in forward direction the 
momentum transfer decreases with increasing photon energy \cite{BERNSTEIN81}
and $|{\bf q}|< 300$ MeV is sufficient. 
The RPA results for initial oxygen and calcium nuclei
are indicated in Fig.\ref{fig:ox3} and Fig.\ref{fig:ca3} at
$|{\bf q}|=200$MeV.
Consistent with the observation for the unnatural parity states in the 
transverse $K^{*}$ response, the residual interaction 
is rather weak. This can be studied in Fig.~\ref{fig:ox4} which compares
the mean field and RPA predictions for $J=1$. The labels indicate the
particle-hole states on the mean field level. 
Because of the very weak spin-orbit force felt by the hyperons, 
the states with the $\Lambda$ in the $1p_{1/2}$ or $1p_{3/2}$ are almost 
degenerate and coalesce in the figure.

Although the results for the kaon response are consistent with the observations
for the transverse $K^*$ response, a more unified treatment is desirable.
Going beyond our first contribution to the subject, one must allow
for mixing between spin-longitudinal and spin-transverse modes.
Either on a phenomenological level by introducing a Landau-Migdal parameter, 
or by incorporating higher-order Feynman diagrams,
this will be an important topic for future investigations.
%
% This is file chap5.tex
%%%%%%%%%%%%%%%%%%%%%%%%%%%
%
\section{Conclusions}
\label{summary}
In this paper we studied strangeness-changing response functions as an
alternative approach to hypernuclear structure. In contrast to the
traditional mean-field description, where hypernuclear states are
uncorrelated hyperon-particle---nucleon-hole excitations, they are
treated as any other nuclear excitation that emerges from the response
of normal nuclei to external probes.  From this point of view we
studied hypernuclear spectra using a relativistic random-phase
approximation based on a chiral Lagrangian which successfully
reproduces properties of normal nuclei and where all relevant
meson-baryon vertices are constrained by SU(3)-flavor symmetry.  

The response of the nuclear ground state involves several polarization
insertions describing the propagation of a nucleon-hole and a
$\Lambda$($\Sigma^0$)-particle through the nuclear medium.  To specify
the residual particle-hole interaction we assumed that it is mediated
by the two lightest $S\!=\!-1$ mesons, namely the kaon and its vector
partner the $K^*(892)$.  This lead us to introduce two
strangeness-changing response functions. A vector $K^*$ response and a
pseudo-vector kaon response.  Many-body correlations are included by
iterating the lowest-order response to infinite order.  The primary
difference of the two response functions lies in the structure of
their excitation spectra.  Due to pseudo-vector character of the $YNK$
vertex, the kaon exclusively excites hypernuclear states of unnatural
parity. In contrast, the $K^*$ response contains---with the exception
of the $J^{\pi}\!=\!0^{-}$ state---hypernuclear states of all spin and
parities.

We analyzed hypernuclear spectra for $^{16}$O and $^{40}$Ca nuclei
under the assumption that a nucleon in the ground state is replaced by
a $\Lambda$ particle. For the longitudinal $K^*$ response, which is
primarily determined in the hadronic strangeness-exchange reactions
such as $(K^-,\pi^-)$ and $(\pi^+,K^+)$, the RPA leads to important
corrections over the simple particle-hole predictions.  First, the
repulsive character of the residual interaction decreases the
uncorrelated hypernuclear binding energies considerably. This
generates the characteristic ``quenching and hardening'' of the
response.  Second, we observed important qualitative corrections in
the spectra, as the strong residual interaction mixes single-particles
states of different particle-hole content but of the same spin and
parity.  Consequently, in most cases it is no longer possible to
attribute the excitations to a specific particle-hole transition.
Furthermore, states with nucleons holes in the deepest bound nucleon
shell move into the continuum consistent with the experimental
observation. In contrast to the longitudinal response, correlation
effects are modest in the transverse part of the $K^*$ response.

We also examined the pseudo-vector kaon response responsible for
generating hypernuclear states of unnatural parity. In this first
study of the spin-longitudinal modes we observed small corrections
over the simple particle-hole picture, in analogy to our observations
for the spin-transverse $K^*$ response. In the future we plan to study
this kaonic response in a much more comprehensive way. Our goal will
be to establish a dynamical range for which the kaonic enhancement---a
precursor to the kaon-condensed state---may be observed. To this end
the upcoming nuclear $K^{+}$-photoproduction experiments at TJANF may
be of direct relevance to our quest~\cite{CHW91}. Indeed, by measuring
an enhancement of the photoproduction background, relative to that
from a single-nucleon, these experiments may shed light on the exotic
and novel states of matter speculated to exist at the core of neutron
stars.

\acknowledgements
We are pleased to thank J. R. Shepard for useful discussions and comments.
This work was supported in part by U.S.DOE under Grant Nos.
DE-FG03-93ER-40774, DE-FC05-85ER250000 and DE-FG05-92ER40750.
%
%

%%%%%%%%%%%%%%%%%%%%%%%%%%%%%%%%%%%%%%%%%%%%%%%%
% This is file tab.tex
%%%%%%%%%%%%%%%%%%%%%%%%%%%
%
\begin{table}[hbt]
\caption{Hyperon coupling constants.}
\medskip
\begin{tabular}[b]{|c|c|c|c|}
$g^\omega_\Lambda$ & 7.9125  & $g^\omega_\Sigma$ &  8.2582 \\
$g^\varphi_\Lambda$ & 6.0233 & $g^\varphi_\Sigma$ & 6.0753  \\
$f^\omega_\Lambda$ &  -4.7597 & $f^\omega_\Sigma$ &  11.739 \\
$g^K_{\Lambda p}$ & 0.7939   & $g^K_{\Sigma p}$   &  0.1250    \\
$g^{K^*}_{\Lambda n}$ & -7.3016  & $g^{K^*}_{\Sigma n}$ & 3.8698  \\
$f^{K^*}_{\Lambda n}$ & -17.311   & $f^{K^*}_{\Sigma n}$ & -6.5049   
\end{tabular}
\label{tab:coup}
\end{table}
\begin{table}[hbt]
\caption{Vector and tensor potentials.
$V^0$ and $b^0$ are the time like component of the $\omega$ and the 
$\rho^0$ mean field respectively, and $A^0$ is the Coulomb potential.
The couplings of the nucleons to the electromagnetic field are incorporated 
via the anomalous magnetic moments $\lambda_p=1.7928$ and $\lambda_n=-1.9131$ 
and via $\beta_F=-0.08230$ and $\beta_D=-0.41754$.}
\medskip
\begin{tabular}[b]{|c|c|c|c|c|c|}
$\Sigma^p_s$ &  $-g^\varphi_N\varphi$ & 
$\Sigma^p_0$ & 
$g^\omega_N V^0+eA^0+{1\over2}g^\rho_N b^0 
+{e\over 2M}(\beta_F+{1\over 3}\beta_D)\Delta A^0$ &
$\Sigma^p_T$ & 
$\frac{d}{dr}(f^\omega_N  V^0
+{1\over 2}f^\rho_N  b^0
+e \lambda_p  A^0)$ \\
$\Sigma^n_s$ &  $-g^\varphi_N\varphi$ & 
$\Sigma^n_0$ & 
$g^\omega_N V^0-{1\over2}g^\rho_N b^0 
-{e\over 3M}\beta_D\Delta A^0$ &
$\Sigma^n_T$ & 
$\frac{d}{dr}(f^\omega_N  V^0
-{1\over 2}f^\rho_N  b^0
+e \lambda_n  A^0)$ \\
$\Sigma^\Lambda_s$ &  $-g^\varphi_\Lambda\varphi$ & 
$\Sigma^\Lambda_0$ & 
$g^\omega_\Lambda V^0-{e\over 6M}\beta_D\Delta A^0$ &
$\Sigma^\Lambda_T$ & 
$\frac{d}{dr}(f^\omega_\Lambda  V^0
+e \mu_\Lambda  A^0)$ \\
$\Sigma^\Sigma_s$ &  $-g^\varphi_\Sigma\varphi$ & 
$\Sigma^{\Sigma^0}_0$ & 
$g^\omega_{\Sigma^0} V^0+{e\over 6M}\beta_D\Delta A^0$ &
$\Sigma^{\Sigma^0}_T$ & 
$\frac{d}{dr}(f^\omega_{\Sigma^0}  V^0
+e \mu_{\Sigma^0}  A^0)$
\end{tabular}
\label{tab:pot}
\end{table}
%
%
%%%%%%%%%%%%%%%%%%%%%%%%%
%
\section*{Figures}
\global\firstfigfalse
\newpage
\newpage
\newpage
\newpage
\newpage
\newpage
\newpage
%%
%
%###########################################################################%

\begin{references}
%
\bibitem{ashman88} J. Ashman {\it et. al.,}
                   Phys. Lett. {\bf 206B} (1988) 364; 
                   Nucl. Phys. {\bf B328} (1989) 1. 
%
\bibitem{review98} Two excellent recent reviews on strange-quark 
                   matter are: 
                   Carsten Greiner and J\"urgen Schaffner-Bielich, 
                   {\it ``Physics of Strange Matter''}, 
                   {\tt nucl-th/9801062};
                   Jes Madsen, {\it ``Physics and Astrophysics of 
                   Strange Quark Matter''},
                   {\tt astro-ph/9809032}.
%
\bibitem{BANDO90} H. Band\={o}, T. Motoba and J. \u{Z}ofka,
                  {\em Int. J. of Mod. Phys. A {\rm 5}} (1990) 4021.
%
\bibitem{NAGELS78} M. M. Nagels, T. A. Rijken and J. J. de Swart, 
                   Phys.\ Rev.\ D 17 (1978) 768.\\
                   P. M. M. Maessen, T. A. Rijken and J. J. de Swart,
                   Phys.\ Rev.\ C 40 (1989) 2226.
%
\bibitem{HOLZENKAMP89} B. Holzenkamp, K. Holinde and J. Speth,
                       Nucl.\ Phys.\ A 500 (1989) 485. \\
                       A. Reuber, K. Holinde, H.-C. Kim and J. Speth,
                       Nucl.\ Phys.\ A 608 (1996) 243.
%
\bibitem{SCHAFFNER93} J. Schaffner, C. B. Dover, A. Gal, C. Greiner,
                      and H. St\"ocker,
                      Phys.\ Rev.\ Lett. \ 71 (1993) 1328.
%
\bibitem{PAPAZOGLOU98a} P. Papazoglou, S. Schramm, J. Schaffner-Bielich, 
                       H.St\"ocker and W. Greiner, 
                       Phys.\ Rev.\ C 57 (1998) 2576.
%
\bibitem{MUELLER99} H. M\"uller, Phys. \ Rev. C \ 59 (1999) 1405.
%
\bibitem{RAYET81} M. Rayet, Nucl. \ Phys.\ A 367 (1981) 381.
%
\bibitem {RUFA87} M. Rufa,  H. St\"ocker, J. Maruhn, W. Greiner,
                  and P. -G. Reinhard, 
                  J. \ Phys. G 13 (1987) L143.\\
                  J. Mare\v{s} and J. \v{Z}ofka, Z.\ Phys.\ A 333 (1989) 209.
%
\bibitem{BRUECKNER76} W. Br\"uckner {\em et al.}, Phys. \ Lett. \ 62B
                      (1976) 481.
%
\bibitem{BERNSTEIN81} A. M. Bernstein, T. W. Donnelly and G. N. Epstein,
                   Nucl.\ Phys.\ A 358 (1981) 195c.
%
\bibitem{COHEN89} J. Cohen, {\em Int. J. of Mod. Phys. A {\rm 4}} (1989) 1.
%
\bibitem{BENNHOLD89} C. Bennhold and L. E. Wright, Phys. \ Rev. C \
                     39 (1989) 927.
%
\bibitem{HUEFNER74} J. H\"ufner, S. Y. Lee and H. A. Weidenm\"uller,
                    Nucl. \ Phys. \ A \ 234 (1974) 429.
%
\bibitem{DOVER80} C. B. Dover, L. Ludeking and G. E. Walker,
                  Phys. \ Rev. \ C \ 22 (1980) 2073.
%
\bibitem {FST97} R. J. Furnstahl, B. D. Serot, and H.-B. Tang, Nucl.\ Phys.\
                   A 615  (1997) 441.
%
\bibitem{CHIANG79} H. C. Chiang and J. H\"ufner, Phys. \ Lett. \ 84B (1979)
                   393.
%
\bibitem{HUEFNER74b} J. H\"ufner, S. Y. Lee and H. A. Weidenm\"uller,
                     Phys. \ Lett. \ 49B (1974) 409.
%
\bibitem{PDG} C. Caso {\em et al.}, The \ European \ Physical \ Journal \ C 3 
              1 (1998).
%
\bibitem{MEISSNER97} U.-G. Mei\ss{}ner and S. Steininger,
                     Nucl.\ Phys.\ B 499 (1997) 349.
%
\bibitem{SCHAFFNER94} J. Schaffner, C. B. Dover, A. Gal, C. Greiner,
                      D. J. Millener and H. St\"ocker,
                      Ann.\ Phys.\ (N.Y.) 235 (1994) 35.
%
\bibitem{MUELLER99b} H. M\"uller and J. R. Shepard, {\texttt nucl-th/9907079}
                     (1999).
%
\bibitem{FETWAL71} A. L. Fetter and J. D> Walecka,
                   {\it ``Quantum Theory of Many-Particle Systems'',}
                   (McGraw-Hill, New York, 1971).
%
\bibitem{SERWAL86}  B.D. Serot and J.D. Walecka, Adv. in
                    Nucl. Phys. {\bf 16}, J.W. Negele and E. Vogt, 
                    eds. (Plenum, N.Y. 1986); 
	            Int. Jour. Mod. Phys. E {\bf 6} (1997) 515.
%
\bibitem{HOROWITZ90} C. J. Horowitz and J. Piekarewicz,
                     Nucl.\ Phys.\ A 511 (1990) 461.
%
\bibitem{EDMONDS57} A. R. Edmonds, Angular momentum in quantum mechanics,
                     (Princeton University Press, Princeton, 1957).
%
\bibitem{POVH80} B. Povh, Nucl.\ Phys.\ A 335 (1990) 233.
%
\bibitem{CHRIEN79} R. E. Chrien, {\em et al.}, Phys. \ Lett. 89B
(1979) 31.
%
\bibitem{CHW91}     C. Hyde-Wright,{\it Quasifree Strangeness
                    Production in Nuclei}, Hall B, Experiment Number {\bf
                    E-91-014}.
\end{references}
\end{document}